\documentclass[12pt,amsfonts]{article}  
\usepackage{amsmath}
\def\ts{\textstyle}
\def\t{\textstyle}        

\def\fiveoverfour{\textstyle\frac{5}{4}}

\newcommand{\bea}{\begin{eqnarray}}
\newcommand{\eea}{\end{eqnarray}}
\newcommand{\beq}{\begin{equation}}
\newcommand{\eeq}{\end{equation}}

\newcommand{\tr}{{\rm tr}}

\def\one{1\hskip-.37em 1}

\def\tr{{{\rm {Tr}}}}
\def\pa{\overrightarrow{p}}
\def\Pa{\overrightarrow{P}}
\def\qa{\overrightarrow{q}}
\def\Qa{\overrightarrow{Q}}
\def\Ra{\overrightarrow{R}}
\def\Sa{\overrightarrow{S}}
\def\half{{\textstyle{\frac{1}{2}}}}

\def\quarter{\textstyle{\frac{1}{4}}}

\def\H{{\cal H}}

\def\H{{\cal H}}

\def\l{\lambda}

\def\op{{\overrightarrow p}}
\def\oq{{\overrightarrow q}}
\def\oP{{\overrightarrow P}}
\def\oQ{{\overrightarrow Q}}
\def\ox{{\overrightarrow x}}
\def\oy{{\overrightarrow y}}

\def\oS{{\overrightarrow S}}

\def\ra{\rightarrow}

\def\tint{{\textstyle\int}}

\def\s{\hskip.08em}
\def\d{\partial}

\def\b{\begin{eqnarray}}  
\def\e{\end{eqnarray}}    
\def\bn{\begin{eqnarray}}  
\def\en{\end{eqnarray}}   
\def\<{\langle}
\def\>{\rangle}

\def\no{\nonumber}

\def\{{\lbrace}
\def\}{\rbrace}
\def\of{{\overrightarrow f}}
\def\op{{\overrightarrow p}}
\def\oq{{\overrightarrow q}}
\def\oP{{\overrightarrow P}}
\def\oQ{{\overrightarrow Q}}
\def\ox{{\overrightarrow x}}
\def\oy{{\overrightarrow y}}

\def\oS{{\overrightarrow S}}

\title{Matrix Models and Large-$N$ Behavior}
\author{John R. Klauder\footnote{Email: klauder@phys.ufl.edu}\\
Department of Physics and\\Department of Mathematics\\
University of Florida\\
Gainesville, FL 32611-8440}
\date{ }
\begin{document}
\maketitle
\begin{abstract}
Following the procedures by which ${\bf O}(N)$-invariant real vector models and their large-$N$ behavior have previously been solved,
we extend similar techniques to the study of real symmetric $N\times N$-matrix models with
${\bf O}(N)$-invariant interactions. Proper extensions to $N=\infty$ are also established. While no $1/N$-expansions are involved in our analysis, a brief comparison of our procedures with traditional $1/N$-expansion procedures is given. \newline

 
\end{abstract}
\section{Introduction} In earlier work \cite{RS,RS2} we have developed the quantum theory of real, classical vector  models with Hamiltonians of the form
   \bn H(\op,\oq)=\half\s[\s\op^2+m_0^2\s\oq^2\s]+\l\s(\oq^2)^2 \;,\label{vm} \en
   where $\op=(p_1,p_2,\ldots,p_N)$, $\oq=(q_1,q_2,\ldots,q_N)$, $\op^2=\Sigma_{n=1}^N\s p_n^2$, $\oq^2=\Sigma_{n=1}^N\s q_n^2$; here $N\le\infty$, and when $N=\infty$, we require that $\op^2+\oq^2<\infty$. In addition, $\op$ and $\oq$  denote canonically conjugate variables with a Poisson bracket such that $\{q_a,p_b\}=\delta_{ab}$.
   Also, the choice of a quartic interaction is not special since we are also able to analyze more general potentials of the form $V(\oq^2)$.

   In Sec.~2, we briefly review the quantization process for this model but let us note here that if we assume, as customary, that the solution for $N=\infty$ involves a ground state that is real, unique, and ${\bf O}(\infty)$ invariant, then only trivial ($=$ free) models exist. A proper solution is found outside standard procedures and involves reducible operator representations.

   In this paper, we take up the quantization of real, symmetric matrix models which have classical Hamiltonians of the form
       \bn H(p,q)=\half\s[ \tr(p^2)+m_0^2\s \tr(q^2)\s]+\l\s\tr(q^4)\;,  \label{mm}\en
       where $p\equiv\{ p_{ab}\}_{a,b=1}^{N,N}$, and $p_{ba}=p_{ab}$, $q\equiv\{q_{ab}\}_{a,b=1}^{N,N}$, and $q_{ba}=q_{ab}$, $\tr(p^2)=\Sigma_{a,b=1}^{N,N} p_{ab}\s p_{ba}$. $\tr(q^2)=\Sigma_{a,b=1}^{N,N}  q_{ab}\s q_{ba}$, and $\tr(q^4)=\Sigma_{a,b,c,d=1}^{N,N,N,N} q_{ab}\s q_{bc}\s q_{cd}\s q_{da}$.  The variables $p$ and $q$ are canonically conjugate and obey the Poisson bracket    $\{q_{ab},p_{cd}\}=\half(\delta_{ac}\delta_{bd}+\delta_{ad}\delta_{bc})$.
       Clearly, this Hamiltonian is invariant under transformations where $p\ra O\s p\s\s O^T$, and $q\ra O\s q\s\s O^T$, where $O\in{\bf O}(N)$.
       Other ${\bf O}(N)$-invariant interactions may be considered as well. One example would be a classical Hamiltonian of the form
           \bn H(p,q)=\half\s[ \tr(p^2)+m_0^2\s \tr(q^2)\s]+\l\s[\tr(q^2)]^2\;,\label{vmm}\en
           which shares the same invariance properties, but is actually equivalent to the vector model considered above. Since that vector model required special care in finding a solution, we may expect that some of the matrix models will also require an unconventional analysis.
\subsection{Relation to standard $1/N$-expansion procedures}
             Because models with large symmetry groups, such as ${\bf O}(N)$ and ${\bf U}(N)$ for large $N$, and even $N=\infty$, are frequently considered, it may be useful if we make a brief comparison of our problem to those of the commonly studied $1/N$-expansion program. To begin with, we note that we have initially started with a classical description of the models we shall consider. In that regard, the parameter $N$ only enters in connection with the number of degrees-of-freedom, e.g., $N$ for the real vector models and $N(N+1)/2$ for the real symmetric models. The symbol $N$ is never used as a coefficient in any classical equation, and the symbol $N$ will never appear as a coefficient in any of our quantum equations as well; it is only related to the number of variables.

            In the usual $1/N$-expansion studies, e.g. \cite{1overN,N2,N3,N4}, scant attention is generally paid to any classical version of the model, and importantly, the parameter $N$, which refers to the number of degrees-of-freedom, also figures prominently as a coefficient, a basic factor, in defining both the Hamiltonian operators and the Hilbert space vectors of the quantum models. This difference follows from a focus on planar diagrams associated with the models being studied, and is natural under the circumstances. However, this property also means that although there is a similarity with our studies from the point of view of the number of degrees-of-freedom, the quantum models themselves and even the Hilbert spaces used are fundamentally different in important respects; this is completely so when $N\ra\infty$.

            The motivation behind the present paper is to examine alternative quantization procedures for highly symmetric models with many---eventually, infinitely many---degrees of freedom. Viewed conventionally, such models tend to exhibit divergences when $N\ra\infty$, which raise traditional questions about how to deal with such divergences. We explore alternative quantization procedures and are able to find methods that provide quantization for the vector and matrix models that are divergence free and preserve the role of $N$ simply as a counting parameter for all $N\le\infty$.

          The main difference involved in conventional quantization procedures and our alternative quantization procedures may be readily seen in the following simple example; for additional discussion, see  \cite{eQ,eQ2} and references therein.

      \subsection{Weak correspondence principle}
      In conventional quantization, for a single degree of freedom, suitable canonical  phase-space coordinates $(p,q)$ are promoted to Hermitian operators, $p\ra P$ and $q\ra Q$, where $[\s Q,P\s]=i\hbar\one$, and the classical Hamiltonian $H_c(p,q)$ guides the choice of the Hamiltonian operator $\H=H_c(P,Q)$, modulo terms of order $\hbar$, which is then used in Schr\"odinger's equation. This is the standard classical/quantum connection.

      Our procedures offer a different form of classical/quantum connection.  Recall that the quantum action functional is given by
            \bn A_Q=\tint_0^T \<\psi(t)|\s[\s i\hbar(\d/\d t)- \H\s]\s|\psi(t)\>\,dt\;,\en
            which, under general stationary variations, leads to Schr\"odinger's equation. However, if we limit ourselves to simple translations and simple uniform motion, rather than general variations---which, thanks to Galilean covariance, we can perform as macroscopic observers without disturbing the system---then the states in question are limited to $|\psi(t)\>\ra|p(t),q(t)\>$.  Here, $|p,q\>$ are canonical coherent states, given by
              \bn |p,q\>=e^{\t -iqP/\hbar}\,e^{\t ip\s Q/\hbar}\s|\eta\>\;,\en
              where  $P$ and $Q$ need to be self adjoint to generate unitary transformations, and in the present case, the normalized fiducial vector $|\eta\>$ may be chosen as $|0\>$, an oscillator ground state that satisfies ($Q+iP)\s|0\>=0$. Thus, the restricted form of the quantum action functional becomes
                \bn A_{Q(R)}\hskip-1.3em&&=\tint_0^T \<p(t),q(t)|\s[\s i\hbar(\d/\d t)- \H\s]\s|p(t),q(t)\>\,dt\no\\
                &&=\tint_0^T[\s p(t)\s{\dot q}(t)-H(p(t),q(t))\s]\,dt\;,\en
                namely the classical action functional enhanced by the fact that $\hbar>0$ still.  The Hamiltonian connection, called the {\it Weak Correspondence Principle} \cite{wcp}, is given, assuming that $P$ and $Q$ are irreducible, by
                  \bn H(p,q)\hskip-1.3em&&\equiv\<p,q|\s\H(P,Q)\s|p,q\>\no\\
                           &&=\<0|\s \H(P+p\one,Q+q\one)\s|0\>\no\\
                           &&= \H(p,q)+{\cal O}(\hbar;p,q)\;. \en

                While it is clear, in our approach, that the relation of the classical variables, $(p,q)$, to the quantum variables, $(P,Q)$, is entirely different from the usual procedure, {\it it is of primary importance to appreciate that the new procedures have---at this point---led to the same result as conventional canonical quantization, specifically, the same Hamiltonian operator, modulo terms of order $\hbar$, that is used in the standard approach.}.

               Although there is much more to this alternative version of the classical/quantum connection, including canonical coordinate transformations and Cartesian coordinates, etc., (see \cite{eQ}), the few equations presented here illustrate the alternative procedure that we shall employ in analyzing the vector and matrix models.
\section{Rotationally-Symmetric Vector Models}
 Let us consider the quantization of the vector model with a classical Hamiltonian given by
             \bn H_c( \pa,\qa)=\half\s[\s\pa^2+m_0^2\s\qa^2\s]+\l\s(\qa^2)^2 \;;\en
             a definition of the symbols appears below Eq.~(\ref{vm}).
             As a consequence of rotational invariance, every classical solution is equivalent to a solution for $N=1$ if $\pa\s\|\qa$
             at time $t=0$, or to a solution for $N=2$ if $\pa\!\not\!\|\s \qa$ at time $t=0$. Moreover, solutions for $N=\infty$ may be derived from those for $N<\infty$ by the limit $N\ra\infty$, provided we maintain $(\pa^2+\qa^2)<\infty$.

            A conventional canonical quantization begins with $\pa\ra\Pa$, $\qa\ra\Qa$, which are irreducible operators that obey  $[Q_l,P_n]=i\hbar\delta_{ln}\one$ as the only non-vanishing commutation relation. For a free model, with mass $m$ and $\l=0$, the quantum Hamiltonian
            $\H_0=\half
            :( \Pa^2+m^2\s\Qa^2):$, where $:(\cdot):$ denotes normal ordering with respect to the ground state of the Hamiltonian.  
            It has the feature that the Hamiltonian operator for $N=\infty$ is obtained as the limit of those for which $N<\infty$.
            Moreover, with the normalized ground state $|0\>$ of the Hamiltonian operator chosen as the fiducial vector for canonical coherent states, i.e.,  \bn |\pa,\qa\>=\exp[-i\qa\cdot\Pa/\hbar]\s\exp[i\pa\cdot\Qa/\hbar]\s|0\>\;,\en
            it follows that
              \bn H_0(\pa,\qa)\hskip-1.3em&&=\<\pa,\qa|\s\half:(\Pa^2+m^2\s\Qa^2):|\pa,\qa\>\no\\
                &&=\half\s(\pa^2+m^2\s\qa^2)\en
              as desired, for all $N\le\infty$, provided that $(\op^2+\oq^2)<\infty$.

              For comparison with later relations, we recall the characteristic function (i.e., the Fourier transform) of the ground-state distribution for the free vector model with mass $m$ given by
                  \bn C_0(\of)\hskip-1.3em&&=M^2_0\int e^{\t i\of\cdot\ox/\hbar}\,e^{\t -m\s\ox^2/\hbar}\,\Pi_{n=1}^N\s dx_n\no\\
                    &&=e^{\t-\of^2/4\s m\s\hbar}\;, \en
               where $M_0$ is a normalization factor, and this result formally holds for $N\le\infty$.

              However,  canonical quantization of the interacting vector models with $\l>0$  leads to trivial results for $N=\infty$. To show this, we assume that the Schr\"odinger representation of the ground state of an interacting model is real, unique, and rotationally invariant. As a consequence, the characteristic function of any 
              ground-state distribution has the form (note: $\of^2\equiv|f|^2\equiv\Sigma_{n=1}^N\s f_n^2$ and $r^2\equiv\Sigma_{n=1}^N\s x_n^2$)
                \bn C_N(\overrightarrow{f})\hskip-1.3em&&=\int e^{\t i\Sigma_{n=1}^N f_n\s x_n/\hbar}\,\Psi_0(r)^2\,\Pi_{n=1}^N\s dx_n\no\\
                  &&=\int e^{\t i|f|\s r\s\cos(\theta)/\hbar}\,\Psi_0(r)^2\,r^{N-1}\, dr\s\sin(\theta)^{N-2}\, d\theta\,d\Omega_{N-2}\no\\
                  &&\simeq M'\int e^{\t-\of^2\s r^2/2(N-2)\hbar^2}\,\Psi_0(r)^2\,r^{N-1}\, dr\s d\Omega_{N-2}\no\\
                  &&\ra\int_0^\infty e^{\t-b\s \of^2/\hbar}\,w(b)\,db \en
                  assuming convergence, where a steepest descent integral has been performed for $\theta$, and in the last line we have taken the limit $N\ra\infty$; additionally,
                  $w(b)\ge0$, and $\int_0^\infty w(b)\s db=1$. This is the result based on symmetry \cite{sh}. Uniqueness of the ground state then ensures that $w(b)=\delta(b-1/4{\widetilde m})$, for some $\widetilde{m}>0$, implying that the quantum theory is that of a free theory, i.e., the quantum theory is trivial! In addition, the classical limit of the resultant quantum theory is a free classical theory, which differs from the original, nonlinear classical theory.
                  This kind of result signals an unsuccessful quantization.
\subsection{Nontrivial quantization of vector models}
                  The way around this unsatisfactory result is to let the representations of $\Pa$ and $\Qa$ be {\it reducible}. The weak correspondence principle, namely
                  $ H(\pa,\qa)\equiv \<\pa,\qa|\s\H\s|\pa,\qa\>$, ensures that the enhanced classical Hamiltonian depends only on the proper variables, but, unlike conventional quantization procedures, there is no rule forbidding the Hamiltonian from being a function of other, non-trivial operators that commute with $\oP$ and $\oQ$, thus making the usual operators reducible. A detailed study \cite{RS} of the proper reducible representation, still in accord with the argument above that limits the ground-state functional form to be a Gaussian, leads to the following formulation. Let $\Ra$ and $\Sa$ represent a new set of canonical operators, independent of the operators $\Pa$ and $\Qa$, and which obey the commutation relation $[S_l,R_n]=\i\hbar\delta_{ln}\one$. We introduce two sets of compatible annihilation operators that annihilate the unit vector $|0,0;\zeta\>$, namely
                       \bn &&
                       [\s m(\s\Qa+\zeta\s \Sa)+i\Pa\s]\s|0,0;\zeta\>=0\;,\no\\
                      && [\s m(\s \Sa+\zeta\s \Qa)+i\Ra\s]\s|0,0;\zeta\>=0\;. \en
                        If we introduce eigenvectors $|\ox,\oy\>$ such that $\oQ|\ox,\oy\>=\ox|\ox,\oy\>$ and $\oS|\ox,\oy\>=\oy|\ox,\oy\>$, it follows that
   \bn \<\ox,\oy|0,0;\zeta\>=M\s \exp[-(\ox^2+2\s\zeta\ox\cdot\oy+\oy^2)/2\hbar]\;\;, \label{gs}\en
                        where $M$ is a normalization factor, and the condition $0<\zeta<1$ ensures normalizability. These two annihilation operators lead directly to
                  two related free Hamiltonian operators,
                    \bn &&\H_{0\,PQ}\equiv\half:(\Pa^2+m^2(\Qa+\zeta\Sa)^2\s):\;,\no\\
                        &&\H_{0\,RS}\equiv\half:(\Ra^2+m^2(\Sa+\zeta\Qa)^2\s):\;,\en
                   for which (\ref{gs}) is a common, unique, Gaussian ground state also used to define normal ordering.  
                   Let new coherent states, which span the Hilbert space of interest, be defined with this ground state as the fiducial vector, as given by
                   \bn|\pa,\qa\>\equiv \exp[-i\qa\cdot\Pa/\hbar]\s\exp[i\pa\cdot\Qa/\hbar]\s|0,0;\zeta\>\;,\en
                   and it follows that
                   \bn H(\pa,\qa)\hskip-1.3em&&=\<\pa,\qa|\s\{\s \H_{0\,PQ}+\H_{0\,RS}+4v\s:\H_{0\,RS}^2:\s\}\s|\pa,\qa\>\no\\
                       &&=\half[\pa^2+m^2(1+\zeta^2)\s\qa^2]+v\s\zeta^4\s m^4\s (\qa^2)^2\no\\
                       &&\equiv \half[\s\pa^2+m_0^2\s \qa^2\s]+\l\s(\qa^2)^2\en
                       as required; and this solution is valid for {\it all} $N$, $1\le N\le\infty$,
                       provided that $(\pa^2+\qa^2)<\infty$.
\section{Rotationally-Symmetric Matrix Models}
As discussed in Sec.~1, we now turn our attention to the quantization of classical systems that have a classical Hamiltonian given by
   \bn H_c(p,q)=\half\s[ \tr(p^2)+m_0^2\s \tr(q^2)\s]+\l\s\tr(q^4)\;,\en
   where the variables $p$ and $q$ are $N\times N$ real, symmetric matrices, as explained after Eq.~(\ref{mm}).
   Such models are invariant under matrix transformations $p\ra O\s p\s\s O^T$ and $q\ra O\s q \s O^T$, where
   $O\in {\bf O}(N)$. The free model, with mass $m$ and $\l=0$, is readily quantized by promoting $p\ra P$ and $q\ra Q$, which are Hermitian, symmetric (in their indices)  matrix operators with the property that the only nonvanishing commutator is
 \bn [\s Q_{ab}, P_{cd}\s] =i\hbar \s\half(\s \delta_{ac}\delta_{bd}+\delta_{ad}\delta_{bc}\s)\s\one\;.\en
 The free Hamiltonian operator is given by
    \bn \H_0\equiv\half\s:\s[\s\tr(P^2)+m^2\s\tr(Q^2)\s]:\;, \en
    and the normalized ground state $|0\>$ of this Hamiltonian is  unique and rotationally invariant, and is given by
     \bn \<x|0\>=M'\s \exp[-m\tr(x^2)/2\hbar\s]\;, \en
     where $M'$ is a normalization factor, $x$ is a real, symmetric $N\times N$ matrix, and $|x\>$ are eigenvectors of $Q$, namely, $Q\s|x\>=x\s|x\>$. Coherent states for this example may be given by
      \bn |p,q\>=e^{\t -i\tr(qP)/\hbar}\, e^{\t i\tr(p\s Q)/\hbar}\,|0\>\;, \en
      and it follows that
        \bn H_0(p,q)\hskip-1.3em&&=\<p,q|\, \half\s:\s[\s\tr(P^2)+m^2\s\tr(Q^2)\s]:\s|p,q\>\no\\
            && =\<0|\, \half\s:\s[\s\tr((P+p\one)^2)+m^2\s\tr((Q+q\one)^2)\s]:\s|0\>\no\\
            &&=\half\s[\s\tr(p^2)+m^2\s\tr(q^2)\s]\;, \en
            as desired.

            Again, for comparison purposes, we compute the characteristic function for the ground-state distribution of the free model, which is given by
              \bn C_0(f)\hskip-1.3em&&=M'^2\int\s e^{\t i\tr(f\s x)/\hbar}\,e^{\t-m\tr(x^2)/\hbar}\,\Pi_{a\le b=1}^{N,N}\s dx_{ab}\no\\
              &&=e^{\t-\tr(f^2)/4m\hbar}\;, \en
              where now $f$ denotes a real, symmetric $N\times N$ matrix, and this relation formally holds for all $N\le\infty$.

   When $\l>0$, conventional quantization procedures would require a rescaling of the quartic interaction term by replacing $\l$ by $\l/N$, as dictated by a perturbation analysis. However, under reasonable assumptions, just as in the vector case discussed in Sec.~2, we are able to show that when $\l>0$ and $N=\infty$ the quantum theory for the matrix models is trivial ($=$ free) just as was the case for the vector models. To show this, we again appeal to the ground state which we assume to be real, unique, and rotationally invariant for the models under consideration. While in the vector models rotational invariance meant the ground state was a function of only one variable, namely $\ox^2$, that is not the case for the matrix models. For example, the ground state could be a function of $\tr(x^2)$, $\tr(x^3)$, $\tr(x^4)$, $\det(x)$, etc. We recognize the fact of many invariant forms, but for simplicity we shall  just display only two of them, namely, $\tr(x^2)$ and $\tr(x^4)$.

   As before, we consider the characteristic function of the ground-state distribution in the Schr\"odinger representation given by
   \bn C_N(f)\hskip-1.3em&&=\int e^{\t i\tr(f\s x)/\hbar}\,\Psi[\tr(x^2),\tr(x^4)]^2\,\Pi_{a\le b=1}^{N,N} dx_{ab}\;, \label{int}\label{ee}\en
   where, again, $f$ is a real, symmetric $N\times N$ matrix. As defined, $C_N(f)$ is clearly invariant under rotations such as $f\ra O\s f\s O^T$. As a real, symmetric matrix, we can imagine choosing $O\in{\bf O}(N)$ so as to diagonalize $f$, namely the matrix is now of the form where $f_{ab}=\delta_{ab}\s f_a$, with $f_a$ being the diagonal elements. This means that the expression
      \bn \tr(f\s x)={\ts\sum}_{a=1}^N f_a\s x_{aa}\;, \en
      and thus only the $N$ {\it diagonal elements} of the real, symmetric $N\times N$ matrix $x$
      enter into the exponent of (\ref{ee}).

      Let us introduce a suitable form of spherical coordinates in place of the $N^*\equiv N(N+1)/2$ integration variables $x_{ab}, \,a\le b$. We choose the radius variable $r$ so that  $r^2\equiv\tr(x^2)= \Sigma_{a,b=1}^{N,N}\s x_{ab}^2$, and the first $N$ angles are identified with the diagonal elements of $x$ leading to
       $x_{11} = r \cos(\theta_1)$, and  for $2\le a\le N$, we set
$x_{aa} = r [\s\Pi_{l=1}^{a-1}\sin(\theta_l)\s] \cos(\theta_{a})$; we also extend the latter notation to include $x_{11}$. The remaining $N(N-1)/2$ variables $x_{ab}$ for $a<b$ are expressed in a similar fashion, but they each involve a factor $1/\sqrt{2}$ to account for their double counting in the definition of $r$. In fact, we need not specify the off-diagonal elements in detail as they do not appear in the Fourier exponent term.  Expressed in these spherical variables, the characteristic function is given by
  \bn \hskip-2em C_N(f)\hskip-1.3em&&= K_N\int \exp\{ir\Sigma_{a=1}^N f_a\s\s [\s\Pi_{l=1}^{a-1}\sin(\theta_l)\s] \cos(\theta_{a})/\hbar\}\no\\
  &&\hskip0em\times\Psi[r^2, \tr(x^4)]^2 \,r^{N^*-1}\s dr\,\Pi_{a=1}^N \s[\s \sin(\theta_a)^{N^*-(a+1)}\,d\theta_a\s]\,d\Omega_{N(N-1)/2}\,,\label{yes}\en
  where $K_N$ is the extra coefficient arising from the $2^{-N(N-1)/4}$ factor in the Jacobian. Observe that the multiple $\sin(\theta_a)$ factors in the Jacobian have huge powers with $N^*-(N+1)$ being the smallest of these. As $N\ra\infty$, it follows that these factors all grow (since $N^*=N(N+1)/2$), and thus all such terms are approximately $N^2/2$ for $N\gg1$. That fact forces each $\theta_a$, $1\le a\le N$, to be constrained to an interval of order $1/N$ around $\pi/2$. We can use that fact in a steepest descent evaluation of each of the $\theta_a$ integrals, $1\le a\le N$. In those integrals the factors $\sin(\theta_a)$ in the Fourier exponent may be set equal to unity, and the terms $\cos(\theta_a)$ are effectively linear in their deviation from $\pi/2$. While the argument $\tr(x^2)=r^2$ in the ground state  is independent of any angles, that is not the case for the argument $\tr(x^4)$ and any other rotationally invariant term that may be there. However, unlike the appearance of the factors $\sin(\theta_a)$ in the Jacobian, the appearance of sine or cosine functions of any angles in the ground state most probably do {\it not} enter with enormous powers $O(N^2/2)$---e.g., even for $\det(x)$ the maximum power would be $N$--- and therefore they should have very little influence on the stationary evaluation of the first $N$ angle integrals. On the other hand, even though the terms $\cos(\theta_a)$ in the Fourier exponent are $O(1/N)$, the factors $f_a$ are arbitrarily large and such terms can not be ignored.

  As a consequence, we are led to an approximate evaluation of the integral (\ref{yes}) given by
\bn  C_N(f)\hskip-1.3em&&\simeq K'_N\int \exp\{-r^2\Sigma_{a=1}^N f^2_a\s/2[N^*-(a+1)]\s\hbar^2\}\no\\
    &&\hskip4em \times\Psi[r^2, \tr(x^4)]^2 \,r^{N^*-1}\s dr\,\,d\Omega_{N(N-1)/2}\no\\
      &&\ra \int_0^\infty e^{\t- b\s \tr(f^2)/\hbar}\, W(b)\,db\;, \label{yes2}          \en
       where in the last line, assuming convergence, we have taken the limit $N\ra\infty$; in addition, $W(b)\ge0$, and $\tint_0^\infty W(b)\s db=1$.  This is the result based on symmetry. If we insist on uniqueness of the ground state, then again it follows that $W(b)=\delta(b-1/4\s\overline{m})$, which leads to a trivial (= free) theory for some mass, $\overline{m}$.

\subsection{Nontrivial quantization of matrix models}
    In finding a nontrivial solution for matrix models, we are guided by the procedures used for vector models. For clarity and comparison,  we start with the matrix model (\ref{vmm}) and afterwards consider the matrix model given by (\ref{mm}). Initially, we introduce reducible representations of the basic variables $P$ and $Q$,
    the $N\times N$ Hermitian, symmetric (in their indices) matrix operators, which satisfy Heisenberg's commutation relations. In addition, we introduce a second, and independent, set of similar matrix operators $R$ and $S$.
    We choose a unit vector in Hilbert space, which we call $|0,0;\zeta\>$, $0<\zeta<1$, and require that
      \bn  &&[\s m\s(Q+\zeta S)+i\s P\s]\s|0,0;\zeta\>=0\;,\no\\
            &&[\s m\s(S+\zeta Q)+i\s R\s]\s|0,0;\zeta\> =0\;. \en
            In terms of eigenvectors $|x,y\>$ for both $Q$ and $S$, respectively, it follows that
            \bn \<x,y|0,0;\zeta\>=M\,\exp\{- m\s[\s\tr(x^2+2\s\zeta\s x y+y^2)\s]/2\hbar\s\}\;,\en
            a Gaussian state in accord with the conclusion of (\ref{yes2}).
      There are two related, Hermitian Hamiltonian expressions of interest, given by
        \bn \H_{1\,PQ}\hskip-1.3em&&=\half\s\tr\{[\s m\s(Q+\zeta S)-i\s P\s]\,[\s m\s(Q+\zeta S)+i\s P\s]\}\no\\
            &&=\half\s:\s[\tr(P^2)+m^2\s\tr((Q+\zeta\s S)^2)\s]: \;,\en
            and
           \bn \H_{1\,RS}\hskip-1.3em&&=\half\s\tr\{[\s m\s(S+\zeta Q)-i\s R\s]\,[\s m\s(S+\zeta Q)+i\s R\s]\}\no\\
            &&=\half\s:[\s\tr(R^2)+m^2\s\tr((S+\zeta\s Q)^2)\s]: \;,\en
     and $|0,0;\zeta\>$ is the unique ground state for both of them. For coherent states, we choose
        \bn |p,q\>=\exp[-i\tr(q P)/\hbar]\,\exp[i\tr(p\s\s Q/\hbar)\s]\,|0,0;\zeta\>\;, \en
        and, for the final, total Hamiltonian, it follows that
         \bn H(p,q)\hskip-1.3em&&=\<p,q|\s\{\s H_{1\,PQ}+\H_{1\,RS}+4\s v\s :\H^2_{1\,RS}:\s\}\s|p,q\>\no\\
              &&=\half\s[\s \tr(p^2)+m^2(1+\zeta^2)\s\tr(q^2)\s]+ v\s m^4 \s  \zeta^4 [\s\tr(q^2)\s]^2\no\\
              &&\equiv\half\s[\tr(p^2)+m^2_0\s\tr(q^2)\s]+\l\s[\s\tr(q^2)\s]^2\;,\label{yes3}\en
              as desired.

         The solution associated with Eq.~(\ref{mm}) is different from that just presented, but, again, according to the analysis that led to (\ref{yes2}), we are still obliged to look for a solution based on a Gaussian ground state.
         Once again we start with the canonical matrix operator pair  $P$ and $Q$, as well as the canonical matrix operator pair $R$ and $S$ that we used in the solution of the model (\ref{yes3}). However, we now  need yet another matrix operator pair, namely, the independent canonical matrix pair $T$ and $U$. This time we start with a ground state denoted by the unit vector $|0,0,0;\xi\>$ which is annihilated by three operators, namely
           \bn &&\hskip-2.75em[\s m\s(Q+\xi(S+U))+iP\s]\s|0 , 0,0;\xi\>=0\;,\no\\
               &&[\s m\s(S+\xi\s Q)+iR\s]\s|0,0,0;\xi\>=0\;,\no\\
               &&[\s m\s(U+\xi\s Q)+iT\s]\s|0,0,0;\xi\>=0\;. \en
         If we introduce eigenvectors $|x,y,z\>$ for the three operators $Q, S$, and $U$, respectively, then it follows that in the Schr\"odinger representation, the ground state is given by
          \bn \<x,y,z|0,0,0;\xi\>=M''\exp\{- m\s[\s\tr(x^2+y^2+z^2+2\s\xi(x\s y+ x z))\s]/2\hbar\s\}\;,\en
          and the condition $0<\xi<1/\sqrt{2}$ ensures normalizability.

          There are now {\it four}  Hamiltonian-like expressions of interest, the first three of which have the  vector $|0,0,0;\xi\>$ as their common, unique ground state, namely,
            \bn \H_{2\,PQ}\hskip-1.3em&&=\half\s\tr\s\{\s[\s m\s(Q+\xi(S+U))-iP\s]\,[\s m\s(Q+\xi(S+U))+iP\s]\s\}\no\\
              &&=\half\s:\{\tr\s(P^2)+m^2\s\tr[(Q+\xi\s(S+U))^2]\s\}:\;,\en
             \bn  H_{2\,RS}\hskip-1.3em&&=\half\s\tr\s\{\s[\s m\s(S+\xi\s Q)-iR\s]\,[\s m\s(S+\xi\s Q)+iR\s]\s\}\no\\
              &&=\half\s:\{\tr\s(R^2)+m^2\s\tr[(S+\xi\s Q)^2]\s\}:\;,\en
              \bn  H_{2\,TU}\hskip-1.3em&&=\half\s\tr\s\{\s[\s m\s(U+\xi\s Q)-iT\s]\,[\s m\s(U+\xi\s Q)+iT\s]\s\}\no\\
              &&=\half\s:\{\tr\s(T^2)+m^2\s\tr[(U+\xi\s Q)^2]\s\}:\;,\en
              and the fourth Hamiltonian expression, which also has $|0,0,0;\xi\>$ as a ground state, is given by
              \bn\H_{2\,RSTU}\hskip-1.3em&&=\tr\{[\s m\s(S+\xi\s Q)-iR\s]\,[\s m\s(S+\xi\s Q)+iR\s]\s\no\\
              &&\hskip1.9em\times\s[\s m\s(U+\xi\s Q)-iT\s]\,[\s m\s(U+\xi\s Q)+iT\s]\s\} \no\\
                   &&=\tr\s\{\s:\s[\s R^2+m^2\s(S+\xi\s Q)^2]\s:\,:\s[\s T^2+m^2\s( U+\xi Q)^2]:\}\;.\en
          For coherent states, we choose
           \bn |p,q\>\equiv \exp[-i\tr(qP)/\hbar]\,\exp[i\tr( p\s\s Q)/\hbar]\,|0,0,0;\xi\>\;,\en
           and  it follows, for the final, total Hamiltonian, that
            \bn H(p,q)\hskip-1.3em&&=\<p,q|\s\{ H_{2\,PQ}+\H_{2\,RS}+H_{2\,TU}+v\s\H_{2\,RSTU}\s\}\s|p,q\>\no\\
                    &&=\half\s[\s\tr(p^2)+m^2(1+2\s\xi^2)\s\tr(q^2)\s]+v\s m^4\s\xi^4\s\tr(q^4)\no\\
                    &&\equiv\half\s[\s\tr(p^2)+m^2_0\s\tr(q^2)\s]+\l\s\tr(q^4)\;,\label{yes9}\en
                    as desired! This expression is valid for {\it all} $N\le\infty$.
\subsubsection*{Critical commentary}
       It is worthwhile to examine a natural candidate that was {\it not} chosen for the last model, and  to see why we did not choose that natural ``solution''. We start be asking why was it necessary to employ {\it three} sets of canonical pairs when it seems that {\it two} canonical pairs should be enough. Before focussing on just two operators, however, let us first assume
    that a new unit vector, $|0,0;\xi\>$---hoping for two pairs and not three---is annihilated by three operators, namely
    \bn &&\hskip.03em A\s\s|0,0;\xi\>\equiv(1/\sqrt{2m\hbar})\s [\s m(Q+\xi(S+U))+i\s P\s]\s|0,0;\xi\>=0\;,\no\\
  && B\s\s|0,0;\xi\>\equiv(1/\sqrt{2m\hbar})\s [\s m(S+\xi Q)+i\s R\s]\s|0,0;\xi\>=0\;,\no\\
        && C\s\s|0,0;\xi\>\equiv (1/\sqrt{2m\hbar})\s[\s m(U+\xi Q)+i\s T\s]\s|0,0;\xi\>=0\;. \en
        The matrix annihilation operators $A$, $B$, and $C$ are associated with matrix creation operators, $A^\dag$, $B^\dag$, and $C^\dag$, and
        \bn [\s A_{ab},A^\dag_{cd}\s]=[\s B_{ab},B^\dag_{cd}\s]=[\s C_{ab},C^\dag_{cd\s}]=\half(\delta_{ad}\delta_{cb}+\delta_{da}\delta_{cb})\one\;. \en
        For simplicity hereafter, we sometimes set $m=\hbar=1$. Thus
         $\H_{2\,PQ}=\tr(A^\dag\s A)$, $\H_{2\,RS}=\tr (B^\dag\s B)$, and  $\H_{2\,TU}=\s\tr(C^\dag C)$.
         For a fourth expression, let
          us introduce ${\cal F}\equiv \tr(B^\dag\s B^\dag\s B\s B)$. With coherent states given (with $\hbar$) by
          \bn|p,q\>=\exp[-i\tr(qP)/\hbar]\s \exp[i\tr(p\s\s Q)/\hbar]\s|0,0;\xi\>\;,\en
          it follows (restoring $m$ part way through) that
           \bn W(p,q)\hskip-1.3em&&\equiv \<p,q|\s\{\s \tr(A^\dag A)+\tr(B^\dag B)+4v\s\tr(B^\dag B^\dag B B) \s\}\s|p,q\>\no\\
            &&=\half\s[\s\tr(p^2)+m^2(1+\xi^2)\s\tr(q^2)\s]+v\s m^4\s\xi^4\s\tr(q^4)\no\\
            &&\equiv\half\s[\s\tr(p^2)+m^2_0\s\tr(q^2)\s]+\l\s\tr(q^4)\s\;,\en
            as desired---or so it would seem.

            The evaluation carried out so far is insufficient to determine whether ${\cal F}\equiv
            \tr(B^\dag B^\dag B B)$ is a genuine Hermitian {\it operator} or, instead, merely a {\it form} requiring restrictions
            on both kets {\it and} bras. For ${\cal F}$ to be an acceptable operator, it is necessary that
            \bn \<\phi|\s {\cal F}^\dag\s {\cal F}\s|\phi\><\infty \en
            for a dense set of vectors $|\phi\>$. If we are able to bring the expression ${\cal F}^\dag {\cal F}$ into normal order, without potential divergences, then we can, for example, use coherent states to establish that $
            {\cal F}$ is an acceptable  operator. After a lengthy calculation, it follows that
          \bn   {\cal F}^\dag {\cal F}\hskip-1.3em&& \equiv  \tr(B^\dag B^\dag B B)\, \tr(B^\dag B^\dag B B)\no\\
                  && = \;: \tr(B^\dag B^\dag B B) \,\tr(B^\dag B^\dag B B) : \no\\
     &&\hskip2em + : \tr(B^\dag B^\dag B B B B^\dag ) :
                   + : \tr(B^\dag B^\dag B B B^\dag B) :\no\\
  &&\hskip2em +\,:\tr(B^\dag B^\dag B B^\dag B B) : +\s \tr(B^\dag B^\dag B^\dag B B B) \no\\
  &&\hskip2em  +\,\s(\fiveoverfour+\half\s N\s)\, \tr(B^\dag B^\dag B B) +\, \quarter\,\tr(B^\dag B^\dag) \tr(B B)  \;.  \label{bad}\en
  This calculation was performed for $N\times N$ matrices, and the last line of (\ref{bad}) has a coefficient $N$ which means as $N\ra\infty$ this term would diverge and cause suitable states with two or more excitations to have an infinite expectation value for the supposed operator ${\cal F}=\tr(B^\dag B^\dag B B)$. Hence, ${\cal F}$ is a {\it form} rather than an acceptable operator, and this fact rules out this proposed quantum solution of the model represented by (\ref{mm}). A similar computation shows that $\tr(B^\dag B B^\dag B)$ is also a form for similar reasons, and so is $\tr(B^\dag C^\dag C B)$.

  On the other hand, we now show that $\H\equiv \tr(B^\dag B C^\dag C)\,[\s=\tr(C^\dag C B^\dag B)\s] $ is a genuine operator for all $ N\le\infty\s$!
To do so, let us bring the expression
$\H^\dag \H$ into normal ordered form. It follows that
    \bn \H^\dag\H\hskip-1.3em&&\equiv \tr( C^\dag C B^\dag B)\,\tr( B^\dag B  C^\dag C ) \no\\
     &&=\;:\tr(C^\dag C B^\dag B)\,\tr( B^\dag B C^\dag C): \no\\
     &&\hskip2em +\,\half\s:\tr(C^\dag C B^\dag B C^\dag C): +\,\half\s:\tr(C^\dag C B^\dag C C^\dag B):\no\\
     &&\hskip2em +\,\half\s:\s\tr(C^\dag B B^\dag C B^\dag B):+\,\half\s :\tr( C^\dag C  B^\dag B B^\dag B):\no\\
     &&\hskip2em+\,\quarter\s:\tr(C^\dag C B^\dag B):+\quarter\,:\tr(C^\dag B B^\dag C):\no\\
     &&\hskip2em+\,\quarter\s:\tr(C^\dag B)\,\tr(B^\dag C):+\,\quarter\s:\tr(C^\dag C )\,\tr(B^\dag B):\;,\en
{\it with no factor of $N$}. Thus, for all $N\le\infty$, the expression for $\H$ is an acceptable operator and not merely a form.
And that is why we have chosen to use that operator to build our interaction in (\ref{yes9}).

It should be observed that all the models which were successfully treated have total Hamiltonians that are well defined and do not exhibit infinities even though they deal with nonlinear models and an infinite number of degrees of freedom when $N=\infty$. This property arises because both the free portion of the Hamiltonian and the interaction portion are compatible, genuine operators. 
The spectrum of these total Hamiltonians can be computed, and that exercise has already been partially carried out for the vector model in \cite{RS2}.

\section*{Acknowledgements}
It is a pleasure to thank C.B.~Thorn and E.~Onofri for helpful discussions and/or correspondence regarding traditional $1/N$-expansion techniques. In addition, thanks are extended to J.~Ben Geloun for earlier discussions that centered on alternative matrix models \cite{JBG}, which have led the author to examine the matrix models addressed in the present paper.

\end{document}